\documentclass[aip,apl,twocolumn,superscriptaddress]{revtex4}
\usepackage{hyperref}
\usepackage[latin1]{inputenc}
\usepackage{natbib}
\usepackage{graphicx}
\usepackage{amsmath}

\begin{document}

\title{Spin-Transfer Torque Induced Vortex Dynamics in Fe/Ag/Fe Nanopillars}
\author{V. Sluka}
\affiliation{Institute of Solid State Research, Electronic Properties (IFF-9)and Jülich-Aachen Research Alliance, Fundamentals for Future 
Information Technology (JARA-FIT),  Research Center Jülich GmbH, D-52425 Jülich, Germany}
\author{A. K\'akay}
\affiliation{Institute of Solid State Research, Electronic Properties (IFF-9)and Jülich-Aachen Research Alliance, Fundamentals for Future 
Information Technology (JARA-FIT),  Research Center Jülich GmbH, D-52425 Jülich, Germany}
\author{A. M. Deac}
\affiliation{Institute of Solid State Research, Electronic Properties (IFF-9)and Jülich-Aachen Research Alliance, Fundamentals for Future 
Information Technology (JARA-FIT),  Research Center Jülich GmbH, D-52425 Jülich, Germany}
\affiliation{present adress: Laboratory of Nanomagnetism and Spin Dynamics, Ecole Polytechnique F\'{e}d\'{e}rale de Lausanne (EPFL), CH-1015 Lausanne, Switzerland\\ and SwissFEL, Paul Scherrer Institut, CH-5232 Villingen PSI, Switzerland }
\author{D. E. Bürgler}
\affiliation{Institute of Solid State Research, Electronic Properties (IFF-9)and Jülich-Aachen Research Alliance, Fundamentals for Future 
Information Technology (JARA-FIT),  Research Center Jülich GmbH, D-52425 Jülich, Germany}
\author{R. Hertel}
\affiliation{Institute of Solid State Research, Electronic Properties (IFF-9)and Jülich-Aachen Research Alliance, Fundamentals for Future 
Information Technology (JARA-FIT),  Research Center Jülich GmbH, D-52425 Jülich, Germany}
\author{C. M. Schneider}
\affiliation{Institute of Solid State Research, Electronic Properties (IFF-9)and Jülich-Aachen Research Alliance, Fundamentals for Future 
Information Technology (JARA-FIT), Research Center Jülich GmbH, D-52425 Jülich, Germany}

\begin{abstract}
We report experimental and analytical work on spin-transfer torque
induced vortex dynamics in metallic nanopillars with in-plane
magnetized layers. We study nanopillars with a diameter of $\mathrm{150\,nm}$, containing two Fe layers with a thickness of $\mathrm{15\,nm}$ and $\mathrm{30\,nm}$ respectively, separated by a $\mathrm{6\,nm}$ Ag spacer. The sample geometry is such that it allows for the formation of magnetic vortices in the Fe disks. As confirmed by
micromagnetic simulations, we are able to prepare states where one
magnetic layer is homogeneously magnetized while the other contains a
vortex. We experimentally show that in this configuration spin-transfer torque can excite vortex dynamics and analyze their dependence on a magnetic field applied in the sample plane. The center of gyration is continuosly dislocated from the disk center, and the potential changes its shape with field strength.
The latter is reflected in the field dependence of the excitation frequency. In the second part we propose a novel mechanism for the excitation of the gyrotropic mode in nanopillars with a perfectly homogeneously magnetized in-plane polarizing layer. We analytically show that
in this configuration the vortex can absorb energy from the spin-polarized electric current
if the angular spin-transfer efficiency function is asymmetric. This effect is supported by micromagnetic simulations.
\end{abstract}

\maketitle

\section{Introduction}
Since the theoretical prediction \cite{number1,number2} and
experimental demonstration \cite{number3,number4} of spin-transfer torque induced
effects many studies focussed on phenomena such as current induced magnetization dynamics and switching. Spin-transfer devices are promising candidates for future information
technology. Current-induced switching between discrete magnetization states may be employed in non-volatile memory such as the magnetic random access memory. As
spin-polarized currents can propel steady spin precession
\cite{number5}, spin-transfer torque devices are also envisaged to be
used as integrated microwave sources. Therefore, finding highly
tunable spin-transfer nano oscillators (STNOs) is a matter of great
current interest. After the discovery that spin-transfer torque can
also drive the oscillatory motion of a magnetic vortex
\cite{number6}, the gyrotropic vortex motion is considered as a
promising STNO candidate.
The magnetic vortex structure appears at certain
dimensions as the ground state of ferromagnetic disks. The magnetization circulates around a core with magnetization perpendicular to the plane of the disk. The orientation of the magnetization in the core is called the polarization and the sense of rotation of the in-plane circulation constitutes the vortex chirality. This configuration is a consequence of the competition between the dipolar and exchange energies.
Various material combinations have been used in previous studies on
spin-transfer torque. In case of metallic spin valves, for example,
Co/Cu systems are very common. In our study we use a molecular beam
epitaxy-grown system containing Fe as ferromagnetic and Ag as spacer
material [Fe/Ag(001)]. It has been calculated \cite{number7} that the
Fe/Ag(001) interface exhibits a resistance that strongly depends on
the spin direction thus providing for a high spin polarization.
Additionally it has been shown \cite{number8} that an asymmetry in
(spin channel averaged) ferromagnetic element and spacer resistances
leads to an angular asymmetry in giant magnetoresistance (GMR) and
also in spin torque efficiency. Experimental
evidence for the appearance of this feature in Fe/Ag/Fe(001) spin
valves has been provided by Lehndorff et al. \cite{number9}.
Furthermore, it has been shown for the case of a thin
($\mathrm{2\,nm}$) extended polarizer that vortex motion can be
excited in Fe/Ag/Fe(001) pillars \cite{number10}.

In this work both magnetic layers have a finite lateral extension,
which makes micromagnetic simulations of the system more feasible. As will be shown the measurements agree very well with the calculations. After the experimental part we will
address the question how the vortex motion in a pillar with in-plane magnetized
polarizer is excited and propose a novel mechanism.
We show by analytical calculation and micromagnetic simulations that
the angular asymmetry in spin-transfer torque efficiency can lead to energy absorption of the magnetic vortex from the spin-polarized electric current.
\section{Sample Fabrication}
After cleaning and annealing the GaAs(100) substrate, the sequence of
metallic layers Fe ($\mathrm{1\,nm}$) / Ag($\mathrm{150\,nm}$) /
Fe($\mathrm{30\,nm}$) / Ag($\mathrm{6\,nm}$) / Fe($\mathrm{15\,nm}$) /
Au($\mathrm{50\,nm}$) is deposited by molecular beam epitaxy. The
$\mathrm{1\,nm}$ Fe layer serves as a seed layer for the adjacent Ag
buffer layer, which will later on be the bottom electrode. It follows
the spin valve layer system with an Au capping layer on top, which
will also provide a part of the top electrode. All layers grow in a 
singlecrystalline manner \cite{number11}. Both Fe layers have bcc
structure and exhibit cubic magnetocrystalline anisotropy. Next
the areas of the prospective bottom electrodes are covered with
resist by optical lithography. The uncovered material is then removed
with ion beam etching (IBE). Afterwards resist dots of $\mathrm{150\,nm}$ diameter are placed into the $\mathrm{15\times 15\,\mu m^{2}}$ contact area located in the
middle of the previously defined structures by means of electron beam lithography. We employ a HSQ electron beam resist (fOx-12). Successively, the layers are milled down
practically onto the bottom electrode by IBE resulting in pillars with
diameters of about $\mathrm{150\,nm}$. We again use HSQ to insulate the nanopillars and planarize the sample. Additional protection and insulation is provided for by a $\mathrm{50\,nm}$ thick $\mathrm{Si_{3}N_{4}}$ layer which is deposited by means of plasma-enhanced chemical vapor deposition (PECVD). In the next step the sample is covered with optical resist apart from
$\mathrm{10\times 10\, \mu m^{2}}$ contact windows above the
nanopillars, which are opened by IBE so that the pillar tops are
cleared from insulating material. A short reactive ion etching (RIE)
step is additionally performed to selectively remove remaining
HSQ. Finally, the top electrode is deposited by a negative optical
lithography step and subsequent lift-off.

\section{Vortex Dynamics: Experiment and Simulation}

\begin{figure}
\includegraphics[width=8.5cm]{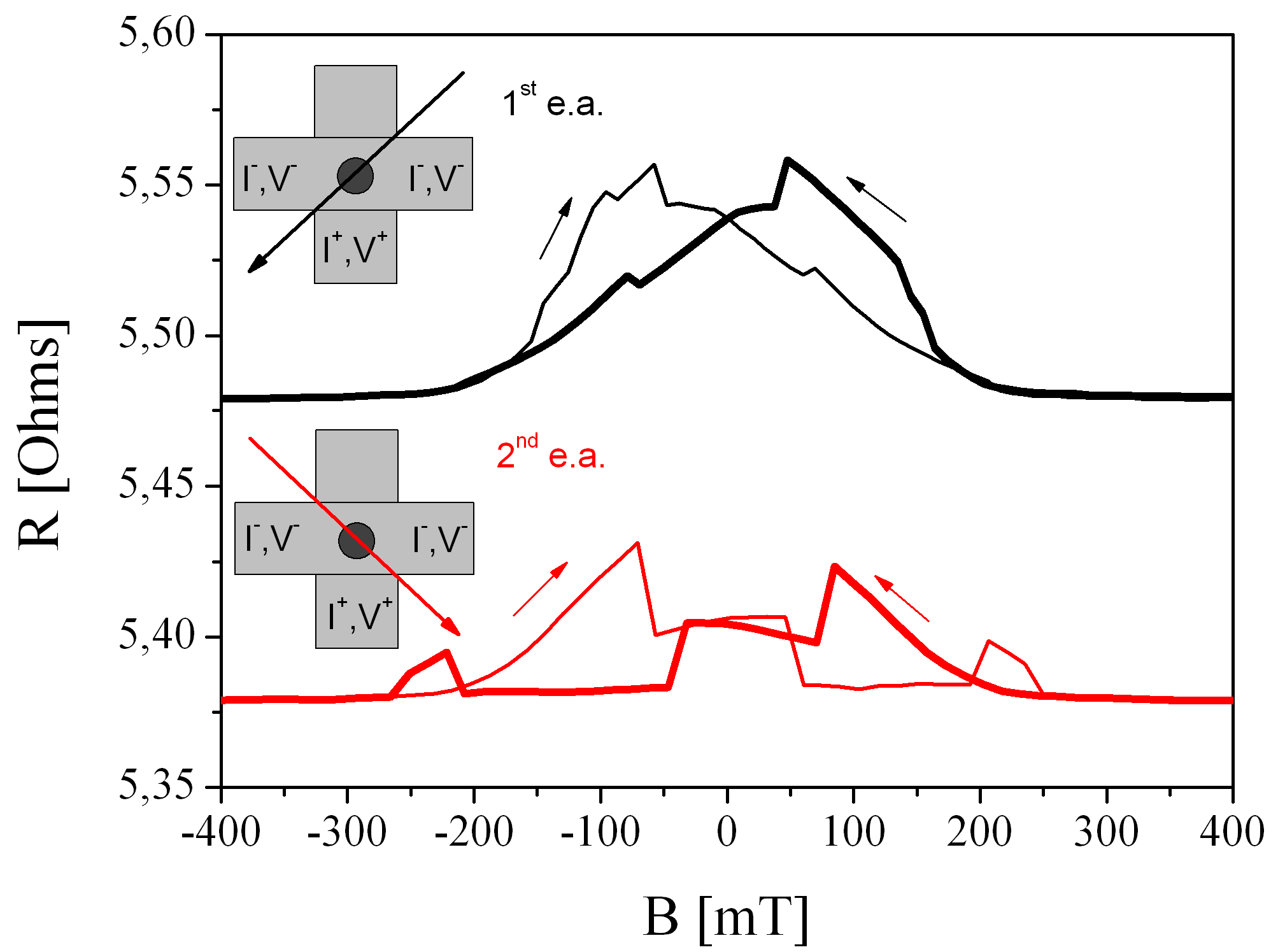}
\caption{\label{fig1}Resistance vs. field dependence for low bias
current ($\mathrm{1\,mA}$). The graphs are vertically shifted for better
visibility. Top (bottom) graph: Applied magnetic field aligned with
easy axis 1 (2). Insets: Schematic top views of the sample, the
arrow indicating the positive field direction. The measurements show
that the two axes are not equivalent.}
\end{figure}
\begin{figure}
\includegraphics[width=8.5cm]{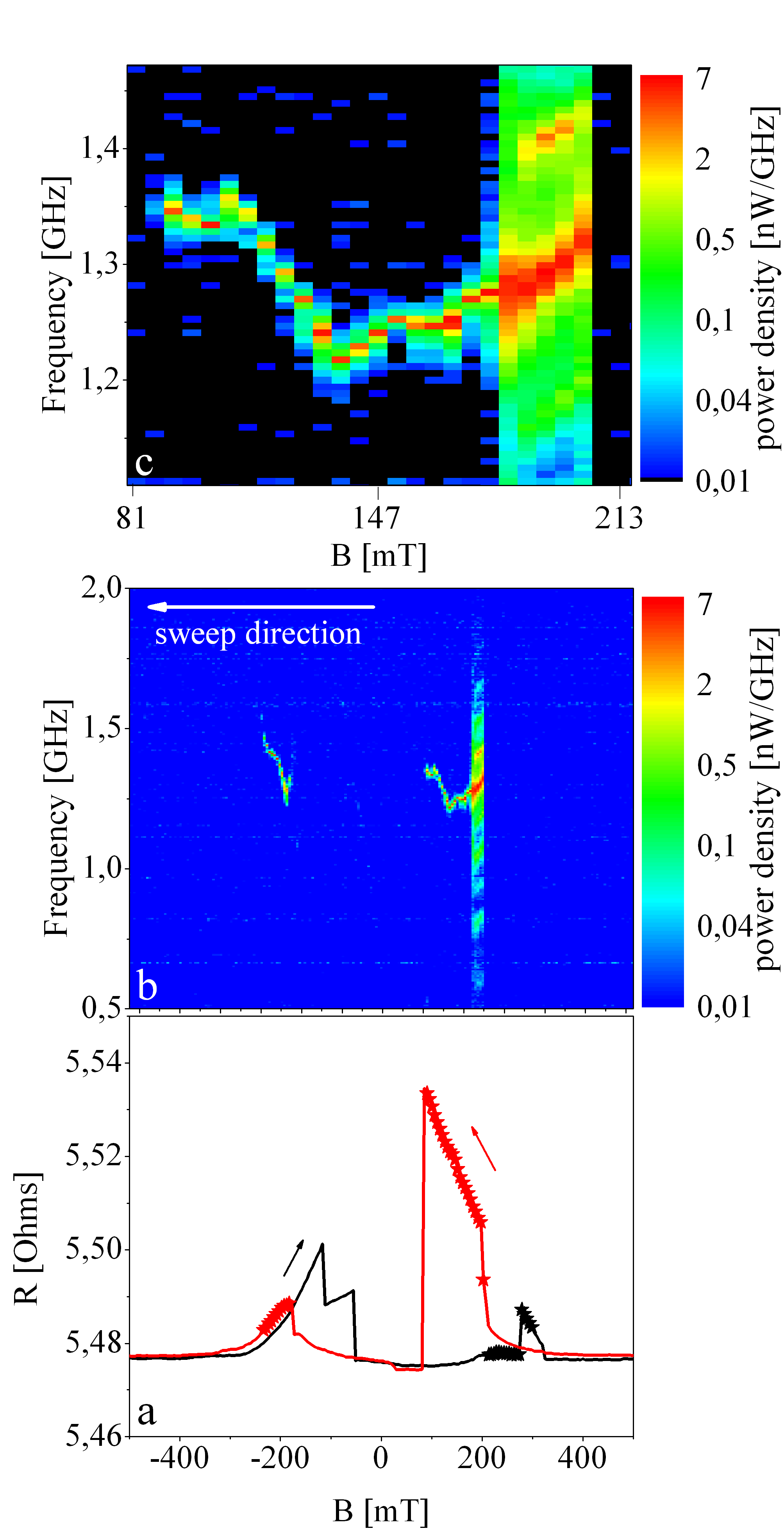}
\caption{\label{fig2}(a) Resistance vs. field dependence for a large bias current of $\mathrm{-17\,mA}$ (electron flow from bottom to top electrode). The field is aligned with easy axis 2. Stars indicate high frequency excitations. (b) Spectra taken at each field value of the field sweep from positive to negative values (red curve in
Fig. \ref{fig2}(a). (c) Enlargement of the vortex
dynamics occurring between $\mathrm{197}$ and $\mathrm{86\,mT}$. 
 }
\end{figure}
\begin{figure}
\includegraphics[width=8.5cm]{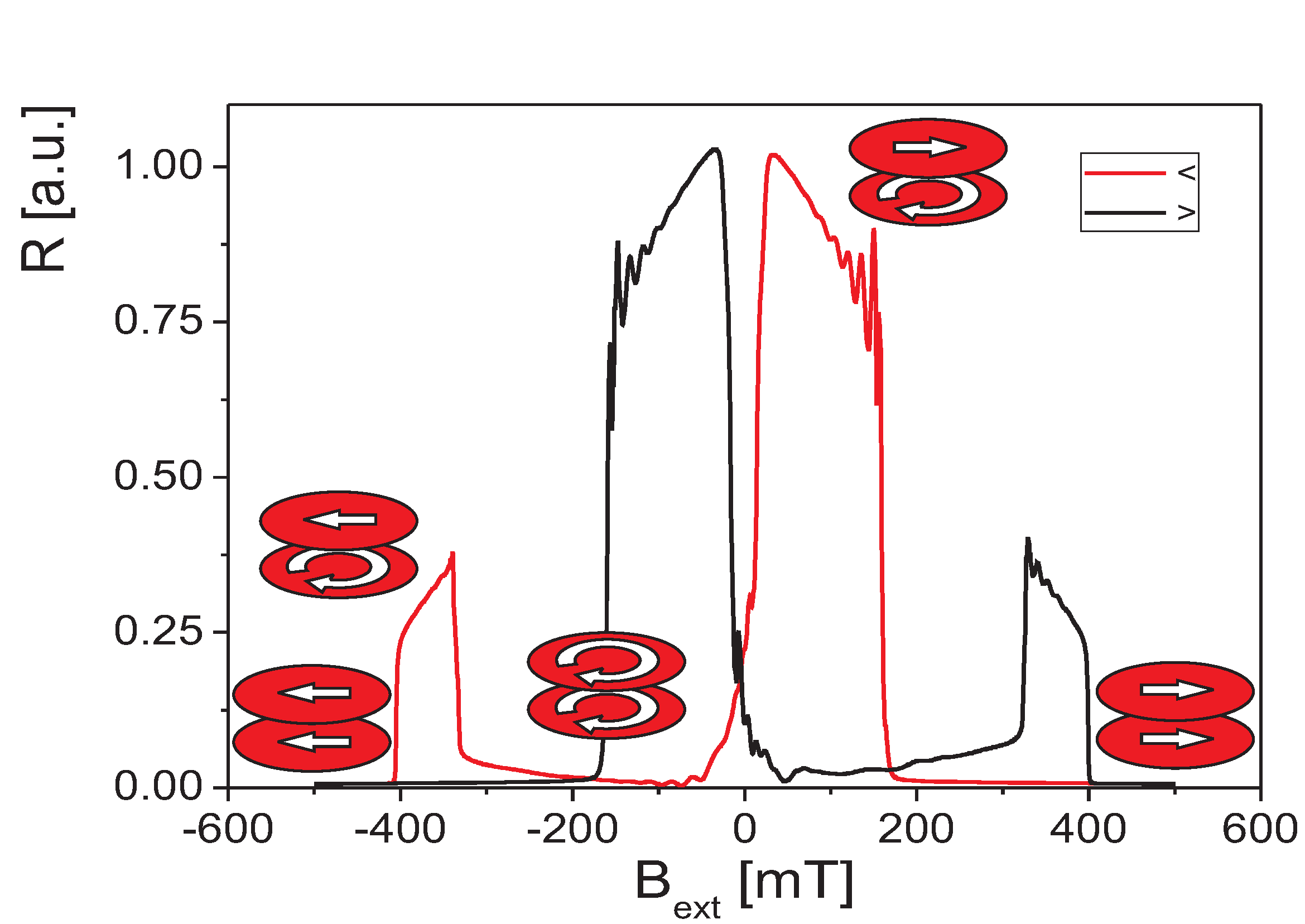}
\caption{\label{fig3} Resistance vs. field dependence obtained from
micromagnetic simulations. The field is aligned with easy axis 2. The
symbols display the associated magnetization states. The simulation
takes into account the Oersted field, which is derived from an
inhomogeneous current density of $\mathrm{-14\,mA}$ total current.
The current distribution reflects the pillar and lead geometry.}
\end{figure}
All measurements are carried out at room temperature. To characterize
the dc behavior of the sample we apply a constant current
while sweeping the magnetic field and measuring the resistance
(2-point measurements). For measurements of the RF voltage we split
the ac and the dc part of the voltage by a bias-T (frequency range: $\mathrm{45\,MHz}$-$\mathrm{26.5\,GHz}$). The signal is then amplified by a $\mathrm{30\,dB}$ amplifier ($\mathrm{0.5}$-$\mathrm{26.5\,GHz}$) and fed into a $\mathrm{50\,GHz}$ spectrum analyzer. The sample holder is home-made and contacts
to the sample leads via spring pins with a diameter of $\mathrm{0.3\,mm}$. The angles between the in-plane easy axes of the cubic anisotropy and the long axes of the top and bottom leads are $\mathrm{45°}$ (insets in Fig. \ref{fig1}). (The two in-plane hard axes of the cubic anisotropy are parallel to the top and bottom leads respectively.)
Fig. \ref{fig1} shows measurements of the dc resistance at low
currents for both easy axes (1st axis: black, 2nd axis: red). The difference of the two cases is likely to be caused by sample imperfections (e.g. electrode contacting, deviations from perfect cylindrical shape of the nanopillar) which can lead to nonsymmetric current distributions and additional contributions to anisotropy. At small applied fields, most interestingly, the axis-2 measurement exhibits states with resistances as low as the saturation resistance. Instead we would expect intermediate up to maximum resistance values due to the dipolar antiferromagnetic coupling between the layers.
To investigate the sample characteristics under high current densities we apply a current of $\mathrm{-17\,mA}$ (Fig. \ref{fig2}a) (electron flow from bottom to top of the pillar). We expect increasing influence of spin torque effects and the Oersted field on the disk magnetizations. The field is applied parallel to easy axis 2 and swept from negative to positive (sweep 1, black) and back to negative values (sweep 2, red). Both sweeps show an increase of the GMR as the applied field is driven from the saturation range towards zero. Once a sufficiently low external field is reached ($\mathrm{-51\,mT}$ for sweep 1, $\mathrm{86\,mT}$ for sweep 2) the sample enters a state of low resistance which persists up to large opposite field values ($\mathrm{274\,mT}$ for sweep 1), where for both sweeps we see abrupt steps in resistance. Afterwards the resistance decreases as the sample is driven into saturation. \\ 
Micromagnetic simulations have been performed with the finite element code called {\tt TetraMag} \cite{number12}.
The magnetization configurations in each disk for the different applied field values are provided by the micromagnetic simulations. Thus a clear physical understanding of the GMR signal is possible.
Typical material parameters of iron, $\mathrm{\mu_{0} M_{s} = 2.15\,T}$ (saturation magnetization), exchange constant $\mathrm{A = 2.1 \times 10^{-11}\,J/m}$ and anisotropy constant $\mathrm{K_{c} = 48\,kJ/m^3}$ are used. The sample volume is discretized into irregular tetrahedrons with cell size of about $\mathrm{1\,nm}$.  The magnetic field is applied along the 2nd easy axis and linearly increased/decreased over a $\mathrm{40\,ns}$ simulation time between the two extrem values. The damping factor is set to $\mathrm{\alpha = 0.1}$.   

The simulations take into account the Oersted field, which is calculated from the
current-density distribution obtained for the nanopillar contact geometry including the leads. The Oersted field distributed in the disk does not have the circular symmetry.
This is due to the asymmetric input-output arrangement of the technical current (see Fig. \ref{fig1}). The maximum field value at the perimeter of the disk is about $\mathrm{41\,mT}$ on one side and $\mathrm{33\,mT}$ on the opposite side.

Fig. \ref{fig3} shows the dependence of the simulated GMR on the applied magnetic field for both sweep directions. The observed dependence is in a good qualitative agreement with the experimentally measured GMR curve for sweep 2. Since the GMR curves are symmetric in the direction of the field sweep we only discuss one sweep direction (from positive to negative).
For field values larger than $\mathrm{200\,mT}$ a low resistance region is found, in which the magnetization in both disks is mostly aligned with the external field.
At about $\mathrm{190\,mT}$, a vortex nucleates in the bottom disk, leading to a jump in the resistance which then further increases linearly with the decrease of the external field. The vorticity of this vortex is determined by the direction of the Oersted field. When the external field approaches zero a vortex with the same vorticity nucleates in the top disk, which is reflected in the low resistance value in this regime. 
With the further decrease of the external field both vortices are moved towards the perimeter of the disks. The vortex of the upper disk is expelled at about $\mathrm{-320\,mT}$ and consequently the resistance increases. The resistance reaches its minimum above $\mathrm{-400\,mT}$ when the vortex of the bottom disk is expelled as well. The parallel orientation of the two magnetizations is restored. The "wiggles" in the resistance accompanying the nucleation of the first vortex are caused by excess energy. In the experiment, this energy is dissipated almost instantly and thus they are not observed.\\

During the measurement depicted in Fig. \ref{fig2}(a) in addition to the dc voltage we measured a spectrum at each field value. The star symbols in the graph mark the presence of peaks in the spectrum representing excitations of the magnetization. Obviously, the single
vortex state is excited by the current. Fig. \ref{fig2}(b) displays
the spectral power density at each field point. In the first field
range from $\mathrm{202}$ to $\mathrm{182\,mT}$ the spectra are broad
($\mathrm{730-1740\,MHz}$) and show multiple peaks of different
amplitudes. As the field is lowered, the frequencies mostly red shift.
At the next field value all secondary peaks have vanished. Starting
from $\mathrm{177\,mT}$ we measure single peak spectra with a peak at
initially $\mathrm{1.275\,GHz}$, which continues the red shift
observed before until $\mathrm{136\,mT}$, where the frequency reaches its minimum of
$\mathrm{1.218\,GHz}$ with a FWHM of $\mathrm{3.2\,MHz}$. Afterwards
the peak frequency increases until $\mathrm{106\,mT}$ to a value of
$\mathrm{1.356\,GHz}$.

If we reverse the current sign the resistance profile
indicates that we again have a vortex in a similar field range, while
the other layer has homogeneous magnetization. But the rich dynamics
observed with the previous current setting do not occur. In order to
rule out thermal activation of the vortex motion we prepare the single
vortex state by applying $\mathrm{-17\,mA}$ and sweeping the applied
field from positive saturation to $\mathrm{151\,mT}$. After reducing
the current stepwise to zero while keeping the field constant we
measure a current loop at this field from $0$ to $\mathrm{-20\,mA}$,
then to $\mathrm{+20\,mA}$ and finally back to $\mathrm{0\,mA}$.  Only
for negative current we see clear excitations. Therefore, thermal
activation cannot explain the excitations, which must be caused by
spin-transfer torque.

These results lead us to the following conclusions: Since both disks
have a geometry, which at low effective fields favors the vortex state,
and due to the match of the resistance profiles obtained by
micromagnetic simulations to the measured ones we have assured that
the investigated state comprises one vortex and one homogeneously magnetized
disk. The homogeneous disk is aligned with the field, therefore its
dipolar field reduces the applied field to sufficiently small
effective values for the vortex to occur in the other disk. No
significant in-plane stray field components are expected to arise from
the vortex, thus the homogeneous disk is subject to the full applied
field. If the observed excitations took place in the homogeneously
magnetized disk, we would expect significantly higher frequencies than
measured.

Another very interesting aspect is the V-shaped dependence of the frequency on the field (Fig. \ref{fig2}(c)).  The most red shifted peak measured occurs at an applied
field of $\mathrm{136\,mT}$.  This value is close to the
micromagnetically computed dipolar field that the spacer facing part
of the vortex disk is exposed to. If we assume that at this point the
dipolar field and the applied field cancel each other, the measured
frequency behavior can most easily be explained: When the vortex
enters the disk the effective field is roughly $\mathrm{60\,mT}$. The
magnetostatic energy minimum, which is the center of the vortex core
motion is shifted towards the disk rim. As the field is lowered the
center of motion moves towards the disk center where the frequency
reaches its minimum value. By further decreasing the applied field
the average vortex position is shifted towards the opposite rim as the
effective field tunes down to about $\mathrm{-50\,mT}$. (At lower
fields the sample enters the double vortex state.) The shifting of the average vortex position is connected to a shape-change of the potential in which the vortex moves. This change of potential shape causes the gyration frequency to first decrease on the way to the disk center and afterwards increase again while the vortex is shifted towards the rim.

We therefore conclude that the vortex is excited and this excitation is caused by spin torque. The vortex moves around an average position, which is shifting through the disk as the magnetic field is swept. The curvature of the potential at its minimum is connected to the restoring force. With changing field, that curvature also changes. This is reflected in the observed V-shaped dependence of the frequency on the field.

\section{Vortex Gyration Under Influence of a Homogeneoulsy Polarized Current}
The excitation of the gyrotropic motion of a vortex by a
spin-polarized current is conventionally described by extending the
Thiele equation with a spin torque term \cite{number13,number14}. However, this approach
denies the possibility to excite a vortex with a current that is
homogeneously in-plane polarized \cite{number15}. Khvalkovskiy et al.
\cite{number15} pointed out that an inhomogeneous polarizer can
supply the vortex with energy. Here, we will propose another
mechanism by showing that even a perfectly homogenous polarizer can
pump energy into the gyrotropic motion of a vortex via spin torque, if
the angular spin torque efficiency function is asymmetric. We start with the Landau-Lifschitz-Gilbert equation with spin torque
\begin{equation}
\frac{\partial\mathbf{m}}{\partial t}=-\gamma\mathbf{m}\times  \mathbf{H}_{\mathbf{eff}}+\alpha \mathbf{m} \times \frac{\partial \mathbf{m}}{\partial t}+\sigma jg(\Lambda,\beta) \mathbf{m}\times(\mathbf{m}\times \mathbf{p}),
\end{equation}
where $\mathbf{m}$ and $\mathbf{p}$ are vector fields of unit length denoting the normalized magnetization and the polarization, respectively. $\alpha$ is the Gilbert damping parameter,  $\gamma$ the gyromagnetic ratio and $j$ the current density. The effective field is defined as
\begin{equation}
\mathbf{H}_{\mathbf{eff}}=-\frac{1}{\mu_{0}M_{s}}\frac{\delta W}{\delta \mathbf{m}},
\end{equation}
where $W$ is the energy density of the disk and $M_{s}$ is the saturation magnetization. The spin torque coefficients are $\sigma=\gamma \hbar/(2\mu_{0}M_{s}eL)$ with $e$ being the modulus of the electron charge and $L$ the disk thickness. The spin-transfer torque efficiency function $g(\beta)$, where $\beta$ is the angle between $\mathbf{p}$ and $\mathbf{m}$ ($\cos(\beta)=\mathbf{m}\cdot \mathbf{p}$) according to Slonczewski \cite{number8} reads
 \begin{equation}
g(\Lambda,\beta)=\frac{P\Lambda}{2(\Lambda \cos^{2}(\beta/2)+\Lambda^{-1}\sin^{2}(\beta/2))}
\end{equation}
 with $P\in[0,1]$ being the degree of current polarization. It is related to the difference in spin dependent resistances $R^{+},\,R^{-}$ as $P=(R^{-}-R^{+})/(R^{+}+R^{-})$. The parameter $\Lambda$ is defined according to $\Lambda=\sqrt{AG(R^{+}+R^{-})/2}$, where $\mathrm{A}$ is the contact cross section and $G$ is the spacer conductance as defined in \cite{number8}. Thus $\Lambda$ describes the mismatch between spacer and ferromagnet resistance and is therefore related to spin accumulation at the nonmagnet-ferromagnet interfaces. Let $\mathbf{p}$ from now on be homogeneous and constant in time. By combining equations (1) and (2) we get the power density absorbed by the disk:
\begin{equation}
\frac{\partial W}{\partial t}=-\frac{\alpha \mu_{0}M_{s}}{\gamma}\left|\frac{\partial \mathbf{m}}{\partial t}\right|^{2}-\frac{ \mu_{0}M_{s}}{\gamma}\sigma j g  \left(\frac{\partial \mathbf{m}}{\partial t}\times \mathbf{m} \right)\cdot \mathbf{p}.
\end{equation}
The total power is obtained by an integration of equation (4) over the disk volume. Let the cylinder axis of the disk be the $z$-axis while the disk is parallel to the $x$-$y$-plane. The direction of $\mathbf{m}$ is then described by the two angles $\varphi(\mathbf{x},t)$ and $\theta(\mathbf{x},t)$ where $\varphi$ is the azimuthal angle ($\varphi=0$ is the $x$-direction) and $\theta$ is the angle between $\mathbf{m}$ and the $z$-axis:
\begin{equation}
\mathbf{m}=\sin(\theta)\cos(\varphi){\hat{\mathbf e}}_{x}+\sin(\theta)\sin(\varphi){\hat{\mathbf e}}_{y}+\cos(\theta){\hat{\mathbf e}}_{z}.
\end{equation}
 We switch to cylindrical disk coordinates (radius $\rho$, azimuth $\chi$, height $z$). In order to carry out the integration we assume that the magnetization distribution obeys the conditions:

\begin{subequations}
\begin{eqnarray}
\varphi(\rho,\chi;a,\chi_{v})=\varphi(\rho,\chi-\chi_{v};a,0)+\chi_{v}
\\
\frac{\partial\varphi}{\partial\chi_{v}}=\frac{\partial\varphi}{\partial\chi_{v}}(\rho,\chi-\chi_{v};a)
\\
\theta=\theta(\rho,\chi-\chi_{v};a)
\\
\frac{\partial\theta}{\partial \chi_{v}}=\frac{\partial\theta}{\partial\chi_{v}}(\rho,\chi-\chi_{v};a),
\end{eqnarray}
\end{subequations}

$a$ and $\chi_{v}$ are the polar coordinates of the vortex core, respectively. It it assumed that the magnetization distribution is independend of the $z$-coordinate. The conditions above state that $\mathbf{m}(a,\chi_{v})$ is related to $\mathbf{m}(a,0)$ by a rotation around the $z$-axis. They are for example fulfilled by the well known double vortex ansatz \cite{number16} (using the notation of \cite{number17})
\begin{equation}
\varphi(\rho,\chi;a,\chi_{v})=h_{a}(\rho,\chi-\chi_{v})+\chi_{v},
\end{equation}

\begin{eqnarray}
h_{a}(\rho,\chi)=\tan^{-1}\left(\frac{\rho \sin\chi}{\rho \cos\chi-a}\right)\\ \nonumber
+\tan^{-1}\left(\frac{\rho \sin\chi}{\rho \cos\chi-R^{2}/a}\right)\pm\frac{\pi}{2},\end{eqnarray}
where $R$ is the disk radius and $\theta$ is some bell shaped function like, for instance according to the ansatz of Thomas and Feldtkeller \cite{number18}
\begin{equation}
\theta=\arccos\left[\pm\exp(-2\kappa^{2} r^{2})\right] 
\end{equation}
with
\[r:=\sqrt{(\rho\cos(\chi-\chi_{v})-a)^{2}+(\rho\sin(\chi-\chi_{v}))^{2}}.\]
In equation (9), the $+$ refers to a positive, the $-$ to a negative core polarity and $\kappa$ is determining the core radius. In order to obtain an expression for the energy the vortex absorbes due to spin torque during one revolution when moving on a constant orbit radius $a$ at constant frequency $\omega$ one has to evaluate the expression
\begin{equation}
E_{\mathbf{STT}}=-\int_{0}^{T} \mathrm{d}t \int_{V}\mathrm{d}x^{3} \,\, \frac{ \mu_{0}M_{s}}{\gamma}\sigma j g  \left(\frac{\partial \mathbf{m}}{\partial t}\times \mathbf{m} \right)\cdot \mathbf{p}.
\end{equation} 
Chosing $\mathbf{p}^{T}=(1,0,0)$ equation (10) can be transformed to

\begin{widetext}
\begin{equation}
E_{\mathbf{STT}}=-L\int_{0}^{2\pi}\frac{\mathrm{d}\chi_{v}}{|\omega|} \int_{A} \mathrm{d} \rho\, \mathrm{d} \chi \,\rho \frac{ \mu_{0}M_{s}}{\gamma}\sigma j
 \tilde{g}\left(\Lambda,\sin\theta\cos\varphi(\rho,\chi+\chi_{v};a,\chi_{v})\right) \left[ \left(\frac{\partial \mathbf{m}}{\partial t}\times \mathbf{m}
\right)_{\chi_{v}=0}\times \mathbf{k}(\chi_{v})\right]_{z}
\end{equation}
\end{widetext}

with $\tilde{g}(\Lambda,\sin\theta\cos\varphi)=\tilde{g}(\Lambda,\cos\beta):=g(\Lambda,\beta)$.
$\mathbf{k}^{T}(\chi_{v})=(\sin\chi_{v},\cos\chi_{v},0)$ is a vector related to the actual vortex position. To proceed further it is useful to express $g$ in terms of the parameter $\xi:=\Lambda^{2}-1$. $\xi=0$ now corresponds to the symmetric case where we perform a Taylor expansion in $\xi$:
\begin{equation}
g(\xi,\beta)=\frac{P}{2}+\frac{1-\cos\beta}{4}P\xi\sum^{\infty}_{n=0}\left(-\frac{1}{2}(\cos\beta+1)\xi\right)^{n}
\end{equation}
This expression only converges for all angles if $-1<\xi<1$. Keeping the first order ($n=0$) term in $\xi$ and inserting it into equation (11) yields:
\begin{equation}
\left.\frac{\mathrm{d}E_{\mathbf{STT}}}{\mathrm{d}\xi}\right|_{\xi=0}=\frac{\pi}{8\left|\omega\right|}\frac{\hbar P j }{e}\int_{A}\mathrm{d}x^{2}  \,\,m_{z}[\mathbf{m}\times  {\dot{\mathbf m}}]_{z}.
\end{equation}
The formula clearly shows that only the core region contributes to the energy uptake. The sign of the integral is independent of core polarity and chirality because vortices of opposite polarity revolve at different sense of rotation. The integrand in equation (13) is sensitive to the core shape. Since moving vortices are accompanied by a dip in the $m_{z}$-magnetization \cite{number19,number20}, we take a magnetization distribution of a moving vortex obtained from an OOMMF-simulation for computation of expression (13). By comparison to the energy dissipated during one period
\[E_{\mathbf{damp}}=\frac{\alpha \mu_{0}M_{s}}{\gamma}\int_{0}^{T} \mathrm{d}t \int_{V}\mathrm{d}x^{3} \,\left|\frac{\partial \mathbf{m}}{\partial t}\right|^{2}\approx\left.\frac{\mathrm{d} E_{\mathbf{STT}}}{\mathrm{d} \xi}\right|_{\xi=0} \xi \]
the critical current for steady gyration is estimated to about $\mathrm{3.36\times 10^{9}\,A/cm^{2}}$ (for $P=0.85$, $\Lambda=1.41$ and $L=30\,\mathrm{nm}$). The vortex absorbes energy for negative $j$ (meaning that electrons flow from the polarizer into the vortex) and positive $\xi$ corresponding to $\Lambda >1$.
\begin{figure}
\includegraphics[width=8.5cm]{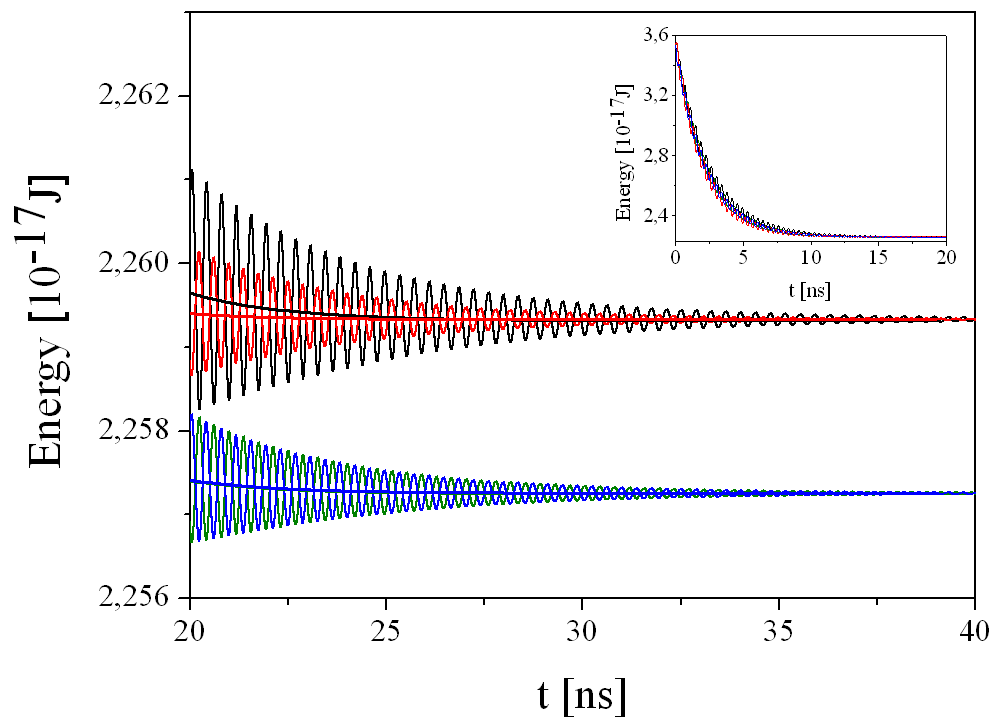}
\caption{\label{fig4}Time dependence of the total energy of the vortex
derived from OOMMF simulations. A vortex initially moving
on an intermediate radius is relaxing while spin
polarized currents of four kinds are applied. Each simulation has the same
initial state. Thin sinusoidal lines: vortex energy. Thick lines:
Fit of exponential decay function $y=A\exp(-x/\tau)+y_{0}$.  Black:
electron flow favors parallel alignment and $\Lambda=1.41$.
$\mathrm{\tau=2.4257\pm 0.0008 \,ns}$.  Red: electron flow favors
antiparallel alignment and $\Lambda=1.41$.  $\mathrm{\tau=2.0326\pm
0.0005 \,ns}$.  Blue: electron flow favors parallel alignment and
$\Lambda=1.0$.  $\mathrm{\tau=2.2236\pm 0.0005 \,ns}$.  Green:
electron flow favors antiparallel alignment and $\Lambda=1.0$.
$\mathrm{\tau=2.2087\pm 0.0005\,ns}$.  Inset: All four curves for
$t<\mathrm{20\,ns}$. }
\end{figure}
 If either the sign of $j$ or $\xi$ is flipped, the term will lead to an additional damping force. The current density estimated above is one order of magnitude higher than usually applied currents. This suggests that the effect is not the only cause for steady gyration. However, simulations clearly confirm it's existence (Fig. \ref{fig4}). We perform four simulations (by OOMMF, lateral cell size $\mathrm{2\,nm}$, one $z$-node, disk diameter $\mathrm{150\,nm}$, disk thickness $\mathrm{30\,nm}$, $\mathrm{\mu_{0}M_{s}=2.14\,T}$, $\mathrm{K_{1}=48\,kJ/m^{3}}$) with the same initial magnetization distribution which is a vortex moving on an orbit with an intermediate radius. The homogeneous polarizer points into the $x$-direction. A current of density $1.5\times 10^{8}\,\mathrm{A/cm^{2}}$ is employed. In the first simulation (black lines) the current sign is chosen such that parallel alignment is favored, while $\Lambda=1.41$ corresponding to $\xi=0.988$. According to equation (13) in first order in $\xi$ the spin current must lower the effective damping in comparison to simulation three (blue), where we have the same current but $\xi=0$. Simulation two (red) is the same as simulation one, but this time the electron flow is reversed so that the current now favors the antiparallel configuration. In that case we expect the strongest damping. Simulation four has the same parameters as three, except that the current sign is negative this time. The damping should be the same as in simulation three. As is shown in Fig. \ref{fig4} all cases qualitatively reproduce the analytical findings: From a fit of an exponential decay function to the data, damping parameters are extracted. Indeed, simulation one exhibits the smallest damping corresponding to the largest time constant $\tau\approx 2.426\,\mathrm{ns}$, simulation two has the largest damping ($\tau\approx 2.033\,\mathrm{ns}$) and simulation three has an intermediate value as expected ($\tau\approx 2.224\,\mathrm{ns}$). Simulation four yields $\mathrm{\tau\approx 2.209\,ns}$ which is very close to the result of three as it should be. In realistic cases $\Lambda$ can be much larger. For the Fe/Ag/Fe(001) system at low temperatures we have an experimental value of $\Lambda=3.4$ obtained from spin-torque measurements leading to $\xi=10.6$ \cite{number9} (and references therein). Equation (12) converges
uniformly for all $\beta\in[0,\pi]$ only if $|\xi|<1$, but was well
suited to calculate the derivative (13). However, it fails if one wants
to compute the energy contribution for larger $|\xi|$. In this case one can expand in $\cos\beta$:
\begin{equation}
g(\Lambda,\beta)=\frac{P\Lambda^{2}}{1+\Lambda^{2}}\sum^{\infty}_{n=0}\left(\frac{1-\Lambda^{2}}{1+\Lambda^{2}}\cos\beta\right)^{n}.
\end{equation}
This series uniformly converges on $\beta\in[0,\pi]$ for every $\Lambda$.

\section{Summary and Discussion}
In the first part we presented measurements of excited states of a nanopillar containing two ferromagnetic disks. One of the disks is in a vortex state while the other is homogeneously magnetized. We deduced that the measured peaks correspond to spin torque driven vortex motion. Several micromagnetic simulations indicate that the vortex resides in the thicker bottom disk. We see the excitations for electron flow from the bottom to the top disk. This would mean that the excitation of the vortex is caused by electrons which are reflected by the top layer. It is not clear why a direct electron flow from the top disk into the vortex cannot excite it. The experiment in \cite{number6} which also has a vortex and a homogeneous polarizer shows excitation only for electron flow into the vortex. On the other hand, the previously reported experiment which also involves the Fe/Ag(001) system \cite{number10} is in accordance with the presented data.  
The V-shaped frequency vs. field dependence of the vortex excitations
has been explained by the in-plane motion of the vortex
potential minimum in the disk and the accompanying shape-change of the potential under the action of the field.

In the second part we have shown that the angular asymmetry in the
spin-transfer torque efficiency function enables the vortex to gain
energy even when the polarizer is perfectly homogeneous. The same
effect can also lead to an additional damping depending on the signs
of current and asymmetry parameter $\xi$. Although the OOMMF
simulations performed so far are too simple to claim capturing the
complex behavior of a thick magnetic disk, they clearly show the
expected qualitative behavior. Nevertheless more advanced simulations
will have to be performed. This novel mechanism is expected to be infuential in the Fe/Ag/Fe(001) system, especially in the experiment reported
by Lehndorff et al. \cite{number10} for two reasons: (i) The
Fe/Ag/Fe(001) system exhibits a rather strong $\xi=10.6$ and (ii) the
polarizing layer is laterally extended and thus much less susceptible
for nonhomogeneous magnetization distributions.\\
\\
A. M. D. acknowledges financial support from the EU project STraDy (MOIF-CT-2006-039772).

\begin{itemize}
\bibitem{number1}
J. Slonczewski, J. Magn. Magn. Mater. \textbf{159}, L1 (1996).
\bibitem{number2}
L. Berger, Phys. Rev. B \textbf{54}, 9353 (1996).
\bibitem{number3}
E. B. Myers, D. C. Ralph, J. A. Katine, R. N. Louie, R. A. Buhrman, Science \textbf{285}, 867 (1999).
\bibitem{number4}
J. A. Katine, F. J. Albert, R. A. Buhrman, E. B. Myers, D. C. Ralph, Phys. Rev. Lett. \textbf{84}, 3149 (2000).
\bibitem{number5}
S. I. Kiselev, J. C. Sankey, I. N. Krivorotov, N. C. Emley, R. J. Schoelkopf, R. A. Buhrmann, D. C. Ralph, Nature (London) \textbf{425}, 380 (2003).
\bibitem{number6}
V. S. Pribiag, I. N. Krivorotov, G. D. Fuchs, P. M. Braganca, O. Ozatay, J. C. Sankey, D. C. Ralph, R. A. Buhrman, Nature Physics \textbf{3}, 489 (2007).
\bibitem{number7}
M. D. Stiles, D. R. Penn, Phys. Rev. B \textbf{61}, 3200 (2000).
\bibitem{number8}
J. Slonczewski, J. Magn. Magn. Mater. \textbf{247}, 324 (2002).
\bibitem{number9}
R. Lehndorff, M. Buchmeier, D. E. Bürgler, A. Kakay, R. Hertel, C. M. Schneider, Phys. Rev. B \textbf{76}, 214420 (2007).
\bibitem{number10}
R. Lehndorff, D. E. Bürgler, S. Gliga, R. Hertel, P. Grünberg, C. M. Schneider, Z. Celinski, Phys. Rev. B \textbf{80}, 054412 (2009).
\bibitem{number11}
H. Dassow, R. Lehndorff, D. E. Bürgler, M. Buchmeier, P. A. Grünberg, C. M. Schneider, A. van der Hart, Appl. Phys. Lett. \textbf{89}, 222511 (2006).
\bibitem{number12}
A. Kakay, E. Westphal, R. Hertel, IEEE Trans. Magn. \textbf{46}, 2303 (2010).
\bibitem{number13}
B. A. Ivanov, C. E. Zaspel, Phys. Rev. Lett. \textbf{99}, 247208 (2007).
\bibitem{number14}
Q. Mistral, M. van Kampen, G. Hrkac, Joo-Von Kim, T. Devolder, P. Crozat, C. Chappert, L. Lagae, T. Schrefl, Phys. Rev. Lett. \textbf{100}, 257201 (2008).
\bibitem{number15}
A. V. Khvalkovskiy, J. Grollier, N. Locatelli, Ya. V. Gorbunov, K. A. Zvezdin, V. Cros, Appl. Phys. Lett.  \textbf{96}, 212507 (2010).
\bibitem{number16}
K. Yu. Guslienko, B. A. Ivanov, V. Novosad, Y. Otani, H. Shima, K. Fukamichi, J. Appl. Phys. \textbf{91}, 8037 (2002).
\bibitem{number17}
A. V. Khvalkovskiy, J. Grollier, A. Dussaux, K. A. Zvezdin, V. Cros, Phys. Rev. B \textbf{80}, 140401 (2009).
\bibitem{number18}
E. Feldtkeller, H. Thomas, Phys. Kondens. Mat. \textbf{4}, 8 (1965).
\bibitem{number19}
M. E. Gouvea, G. M. Wysin, A. R. Bishop, F. G. Mertens, Phys. Rev. B \textbf{39}, 11840 (1989).
\bibitem{number20}
V. Novosad, F. Y. Fradin, P. E. Roy, K. S. Buchanan, K. Yu. Guslienko, S. D. Bader, Phys. Rev. B \textbf{72}, 024455 (2005).
\end{itemize}

\end{document}